\documentclass{PoS}

\title{Hard X-ray surveys and the local AGN population}

\ShortTitle{Hard X-ray surveys and local AGN}

\author{\speaker{Sergey Sazonov}\\
        Space Research Institute, Russian Academy of Sciences\\Moscow, Russia\\
        E-mail: \email{sazonov@iki.rssi.ru}}

\abstract{We discuss some recent achievements in our understanding of
  the local AGN population based on large-area hard X-ray surveys (in
  particular, by the IBIS instrument aboard INTEGRAL), focusing on the
  issue of the fraction of obscured AGN as a function of hard X-ray luminosity.}

\FullConference{11th INTEGRAL Conference Gamma-Ray Astrophysics in Multi-Wavelength Perspective,\\
		10-14 October 2016\\
		Amsterdam, The Netherlands}

\newcommand{\nh}{N_{\rm{H}}}

\begin{document}

\section{Introduction}

Hard X-ray (at energies above $\sim 10$ keV) surveys allow one to
obtain a nearly unbiased census of unobscured and obscured AGN (except
for Compton thick objects, which are relatively difficult to detect
even in hard X-rays). Although there were space-borne hard X-ray
detectors long ago, their sensitivity and/or field of view were not
sufficient to collect good statistics of AGN in the local Universe
(let alone the more distant Universe). The situation changed in 2004,
when the results of the RXTE slew survey were reported
\cite{revetal04}, which had covered the entire extragalactic sky
($|b|>10^\circ$) in the 3--20~keV band with a record sensitivity of
$\sim 10^{-11}$~erg~s$^{-1}$~cm$^{-2}$. Using these data, we
\cite{sazrev04} detected $\sim 100$ AGN and studied a number of key
properties of the local ($z<0.1$) AGN population, namely the
luminosity function and the luminosity dependence of the number ratio
of obscured and unobscured AGN \cite{sazrev04}. Soon thereafter, the
first results of observations by the IBIS and BAT coded-mask
instruments aboard the INTEGRAL and Swift spacecraft, respectively,
appeared. These missions have provided the first sensitive all-sky
surveys at energies above 15~keV and finally opened the era of
extragalactic hard X-ray surveys.

Over the years, hard X-ray source catalogs compiled from IBIS and
BAT data have greatly increased in size and their latest versions
include $\sim 400$ \cite{biretal16,krietal12} and $\sim 700$
\cite{bauetal13} AGN, respectively. This is more than sufficient for
undertaking statistical studies of the local AGN population, provided
that complementary information about the distances, X-ray absorption
column densities and other properties of these objects is
available. Such information has been accumulating over the years
thanks to the X-ray and optical follow-up efforts of several teams, but
usually a few years pass after release of a given hard X-ray catalog
before it achieves high completeness in terms of soft X-ray and
optical information. 

\section{Fraction of obscured AGN as a function of hard X-ray luminosity}

We recently \cite{sazetal15} made use of a highly complete (in terms of
optical identification and X-ray absorption information) sample of 151
non-blazar AGN located at $|b|<5^\circ$, selected in the 17-60 keV
energy band from the INTEGRAL/IBIS 7-year (December 2002--July 2009)
all-sky hard X-ray survey \cite{krietal10}, to investigate if the
declining trend of the fraction of obscured AGN with
increasing luminosity, observed in many studies, is mostly an
intrinsic or selection effect. The sample comprises 67
unobscured (X-ray column density $\nh<10^{22}$~cm$^{-2}$) and 84
obscured ($\nh\ge 10^{22}$~cm$^{-2}$) AGN, including 17 heavily
obscured ($\nh\ge 10^{24}$~cm$^{-2}$) ones. For most of these Compton
thick (or nearly Compton thick) objects, there are now reliable $\nh$
estimates based on high-quality NuSTAR spectroscopic data. 

Using a radiative transfer model based on torus obscuration geometry, we
demonstrated that in addition to a negative bias in finding heavily 
obscured AGN in hard X-ray flux limited surveys there must also exist a
positive bias in detecting unobscured AGN -- due to reflection by the
torus of part of the radiation emitted by the central source toward the
observer. Since the AGN luminosity function
steepens at high luminosities, these observational biases must
inevitably lead to a decreasing observed fraction of obscured AGN with
increasing luminosity even if this fraction has no intrinsic
luminosity dependence. 

\begin{figure}
\includegraphics[width=0.5\textwidth,viewport=0 160 580 710]{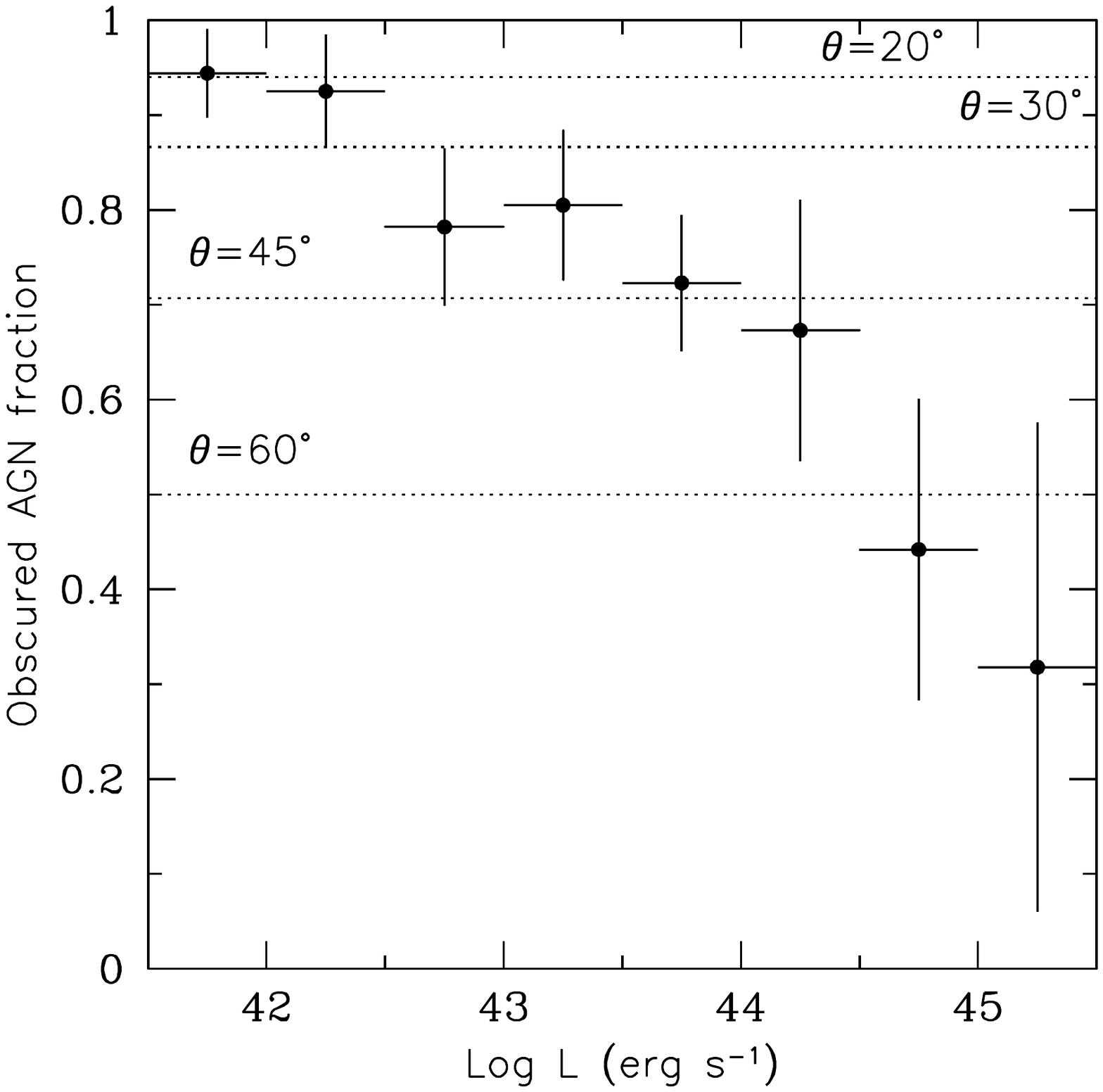}
\includegraphics[width=0.5\textwidth,viewport=0 160 580 710]{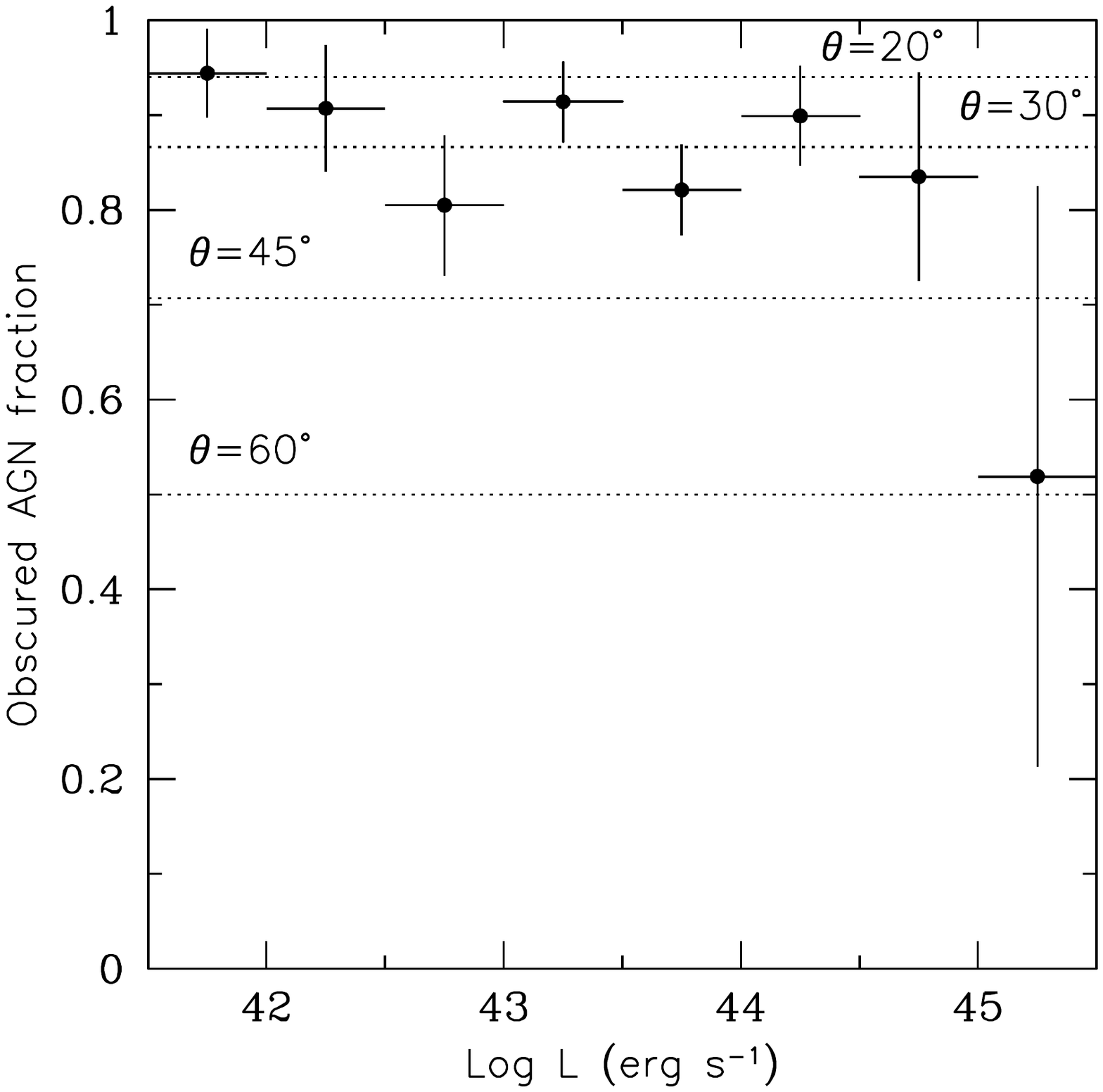}
\caption{\textit{Left:} Reconstructed intrinsic fraction of obscured
  AGN as a function of intrinsic hard X-ray (17--60~keV) luminosity 
  for a torus half-opening angle of $30^\circ$ and an isotropic central
  source. \textit{Right:} The same but for collimated emission
  ($dL/d\Omega\propto\cos\alpha$) from the central source. The dotted
  lines indicate the fraction of the sky obscured from the central
  source by a torus with half-opening angle $\theta=20^\circ$,
  $30^\circ$, $45^\circ$ or $60^\circ$. Adapted from \cite{sazetal15}. 
}
\label{fig:intr_obsc_frac_30}
\end{figure}

We explored two possibilities for the central hard X-ray source in
AGN: (i) isotropic emission (as is assumed in most studies) and (ii)
emission collimated according to Lambert's law,
$dL/d\Omega\propto\cos\alpha$, where $\alpha$ is the angle with
respect to the axis of the torus (which is a fairly natural assumption
in the context of disk-corona geometry). In the former case, the intrinsic 
(i.e. corrected for the biases discussed above) obscured AGN fraction
reconstructed from our sample shows a declining trend with increasing
luminosity, although the inferred intrinsic obscured fraction proves
to be larger than the observed one. Namely, the obscured fraction is
larger than $\sim 85$\% at $L\lesssim 10^{42.5}$~erg~s$^{-1}$ and
decreases to $\lesssim 60$\% at $L\gtrsim 10^{44}$~erg~s$^{-1}$
(17--60~keV). In terms of the half-opening angle $\theta$ of the
torus, this implies that $\theta\lesssim 30^\circ$ in lower luminosity
AGN, and $\theta\gtrsim 45^\circ$ in higher luminosity ones. If,
however, the emission from the central SMBH is collimated, then the
derived intrinsic dependence of the obscured AGN fraction is
consistent with the opening angle of the torus being constant with
luminosity, namely $\theta\sim 30^\circ$ (see
Fig.~\ref{fig:intr_obsc_frac_30}). 

We regard both possiblities -- the intrinsic obscuring AGN
fraction declining with luminosity or being constant -- as feasible,
as the angular emission diagram of the central source in AGN is poorly
known. A more careful comparison of these and
other existing estimates of the ratio of obscured and unobscured AGN
in future work may help get insight into the geometrical and physical
properties of obscuration in AGN, which may be different in X-ray,
optical, infrared and radio bands. Our constraints on
the intrinsic dependence of the obscured AGN fraction on luminosity
can be improved using larger samples of hard X-ray selected AGN from
INTEGRAL, Swift and NuSTAR surveys. However, it will be
practially impossible to improve the current, fairly uncertain
estimate of the obscured AGN fraction at the highest luminosities
($\gtrsim 10^{45}$~erg~s$^{-1}$) in the local Universe, since the
INTEGRAL and Swift all-sky surveys are sensitive enough to detect all
such objects in the local ($z\lesssim 0.2$) Universe and have found
just a few of them.

\begin{figure}
\includegraphics[width=0.5\columnwidth,viewport=0 160 580 710]{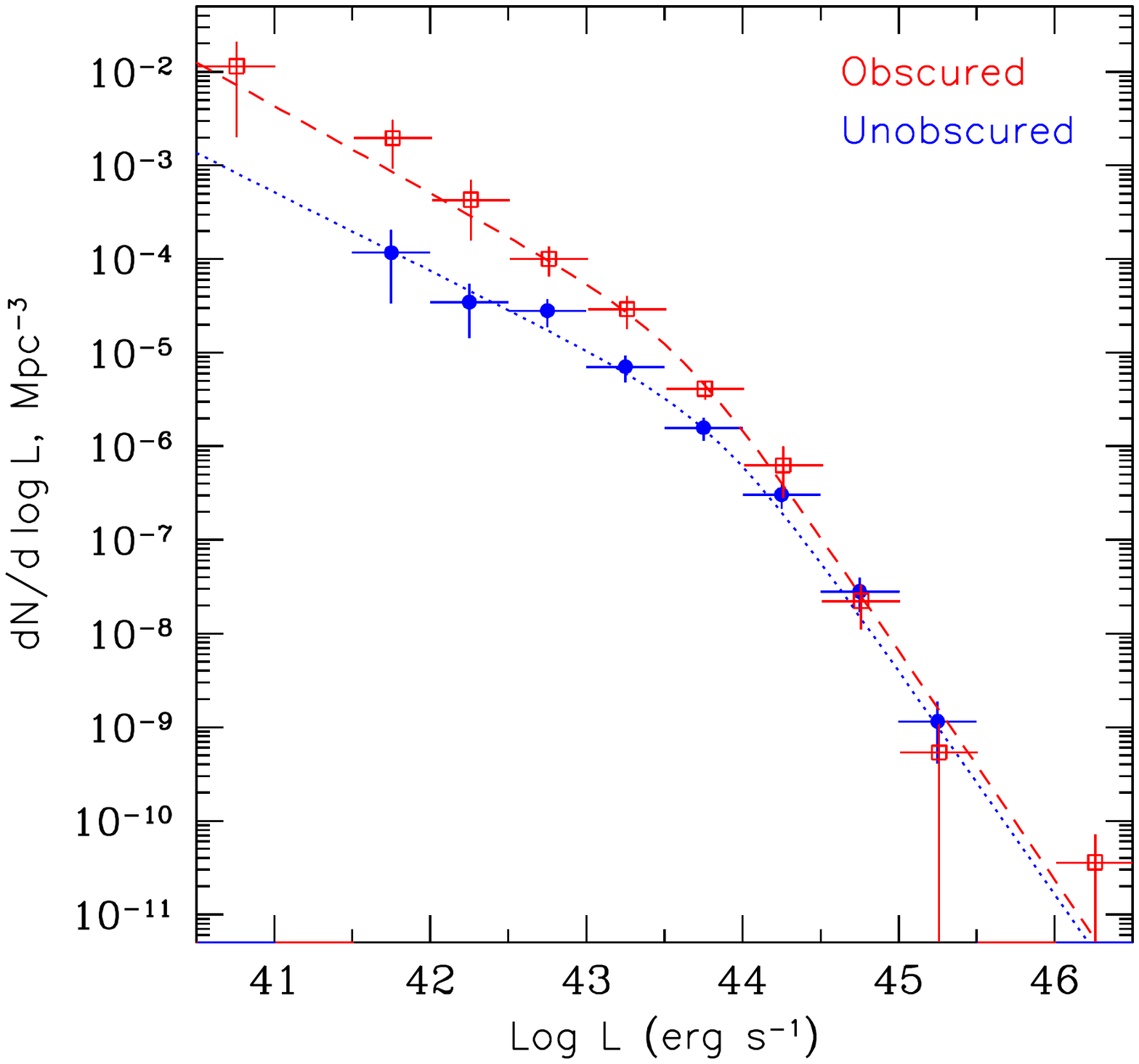}
\includegraphics[width=0.5\columnwidth,viewport=0 160 580 710]{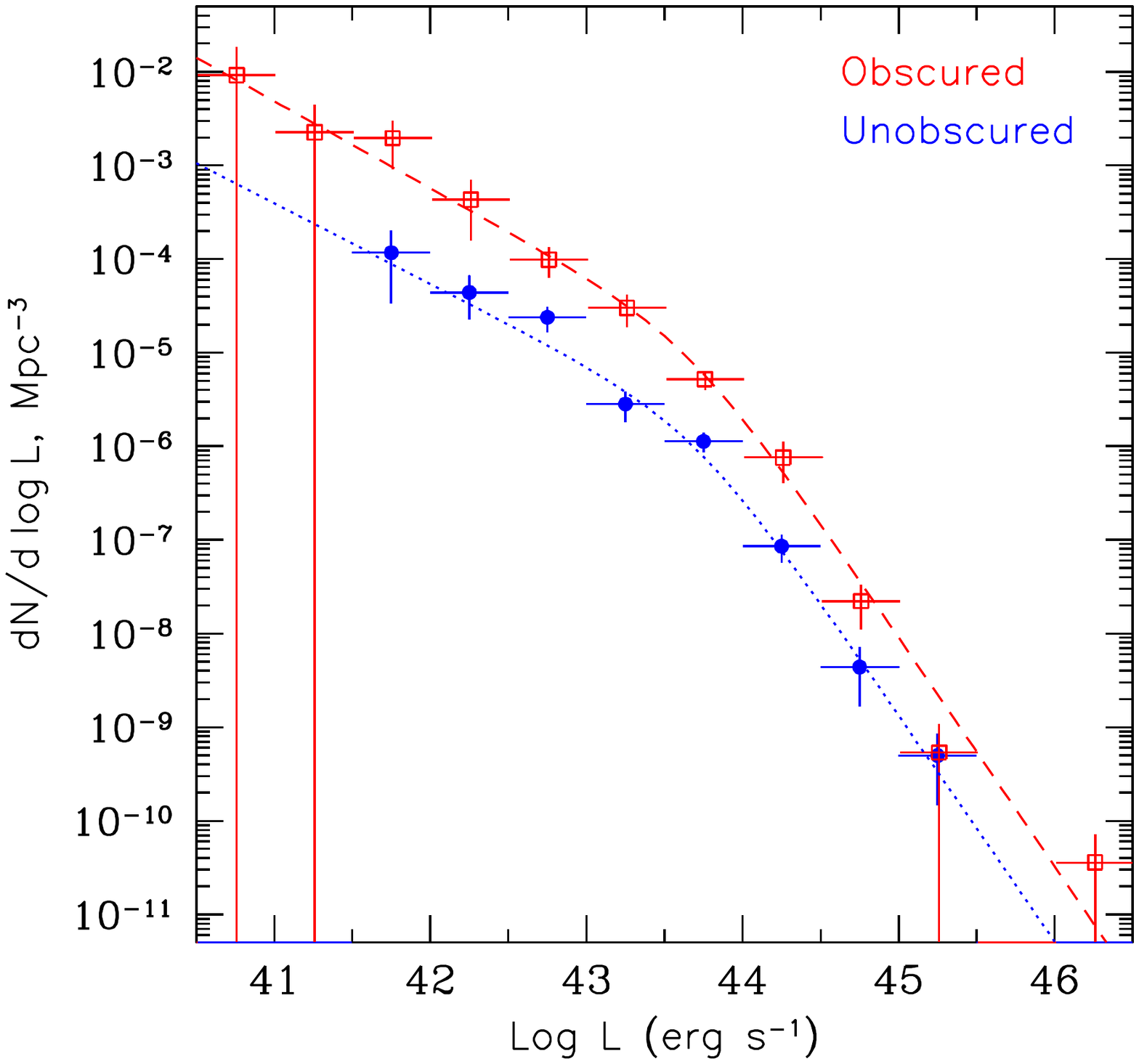} 
\caption{\textit{Left:} Intrinsic hard X-ray (17--60~keV) luminosity
  functions of unobscured (blue filled circles) and obscured (red
  empty squares) AGN, fitted by broken power laws (blue dotted line
  and red dashed line, respectively), reconstructed under the
  assumption of a torus half-opening angle of $30^\circ$ and an
  isotropic central source. \textit{Right:} The same but for the case
  of collimated emission ($dL/d\Omega\propto\cos\alpha$) from the
  central source. Adapted from \cite{sazetal15}.
}
\label{fig:intr_lumfunc_nh}
\end{figure}

We have also reconstructed the intrinsic (corrected for the biases 
discussed above) hard X-ray luminosity functions of local unobscured
and obscured AGN (see Fig.~\ref{fig:intr_lumfunc_nh}) and estimated
the total number density and luminosity density of AGN with
$L>10^{40.5}$~erg~s$^{-1}$ \cite{sazetal15}. These accurate local
measurements may be used as reference $z=0$ values in studying cosmic
AGN evolution and modeling the cosmic X-ray background. 

\section{Deep INTEGRAL extragalactic surveys}

The observational program of INTEGRAL has been mainly dedicated to
Galactic source studies, whereas the high Galactic latitude sky has
been observed less intensively and very inhomogeneously. However, a
number of multi-year campaigns have been performed in the extragalactic
sky, in particular in the 3C~273/Coma region
\cite{paletal08}, around the Large Magellanic Cloud \cite{greetal12}
and most recently around the M81 galaxy
\cite{sazetal14,chuetal14}. Importantly, the sensitivity of the IBIS
instrument in these extragalactic fields continues to grow nearly
proportionally to the square root of exposure showing no significant
contribution of systematic noise. In combination with IBIS large field
of view, this opens up a possibility to collect a significantly large
sample of hard X-ray emitting AGN with fluxes down to a few
$10^{-12}$~erg\,s$^{-1}$ cm$^{-2}$. Such objects, due to their rarity
in the sky, evade NuSTAR deep surveys. 

\begin{figure}
\includegraphics[width=0.47\columnwidth,viewport=0 0 620 620]{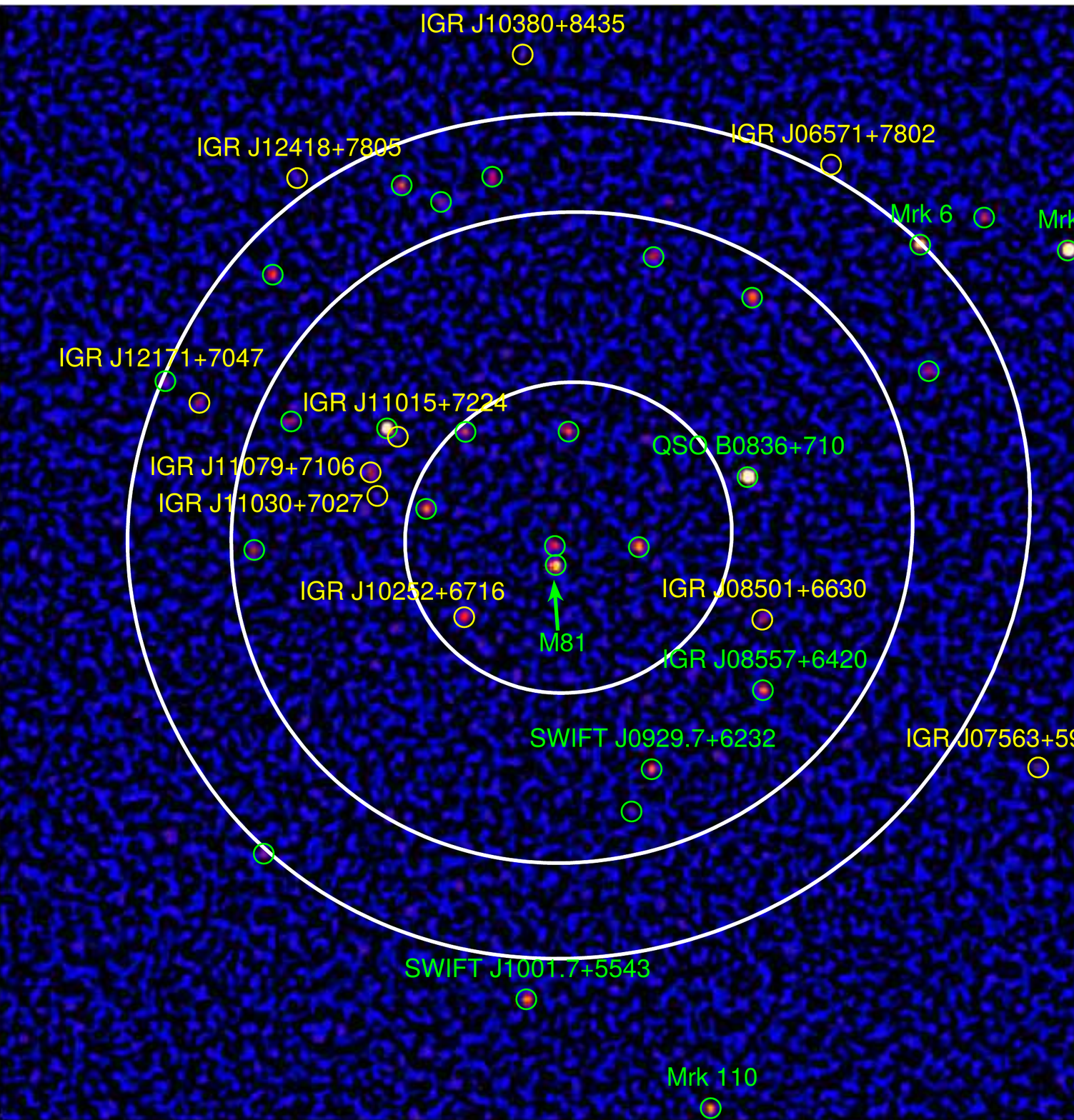} 
\includegraphics[width=0.54\columnwidth,viewport=0 0 480 480]{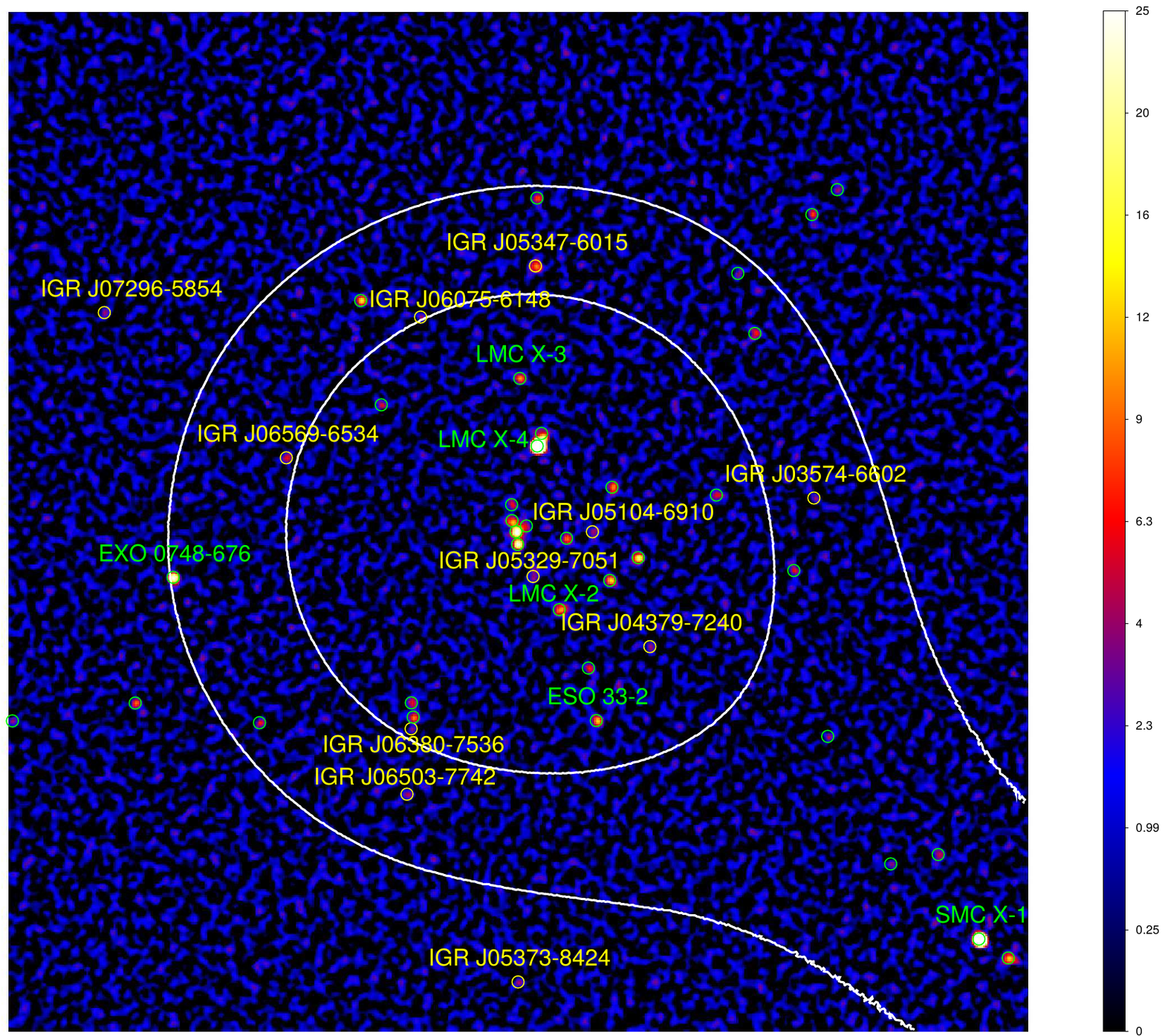}   
\caption{Hard X-ray maps of the M81 and LMC fields, shown in terms of
  significance. Yellow circles denote new sources and green circles
  already known ones. Some of the brightest sources are marked for easy
  navigation. \textit{Left}: The M81 field. The peak exposure 9.7 Ms, 
  contours show exposures of 2, 4 and 8 Ms. \textit{Right:} The LMC
  field. The peak exposure 6.8 Ms, contours  drawn at 2 and 4
  Ms. Adapted from \cite{meretal16}. 
}
\label{fig:maps}
\end{figure}

We have recently presented \cite{meretal16} results of a deep survey
of three extragalactic fields, M81 (dead time corrected exposure of
9.7~Ms), LMC (6.8~Ms) and 3C 273/Coma (9.3~Ms), in the hard X-ray
(17--60~keV) energy band with the IBIS telescope, based on 12 years of
observations (2003--2015). The combined survey reaches a $4\sigma$
peak sensitivity of 0.18~mCrab (2.6$\times$10$^{-12}$ erg s$^{-1}$
cm$^{-2}$) and sensitivity better than 0.25 and 0.87 mCrab over 10\%
and 90\% of its full area of 4900 deg$^{2}$, respectively. We have
detected in total 147 sources at $S/N>4\sigma$, including 37 sources
observed in hard X-rays for the first time (see
Fig.~\ref{fig:maps}). The survey is dominated by AGN (98 identified
objects of this type). The nature of 25 sources remains unknown.

Using these data, we constructed AGN
number-flux relations ($\log{N}$-$\log{S}$) and calculated AGN number
densities in the local Universe for the entire survey and for each of
the three extragalactic fields down to fluxes $\sim3\times10^{-12}$
erg\,s$^{-1}$\,cm$^{-2}$, which is deeper by a factor of two compared
to previous (all-sky) measurements. The AGN number counts for the M81
and 3C 273/Coma fields are consistent with each other, while the LMC
field demonstrates a steeper number-flux distribution (2$\sigma$
deviation from the expected $-3/2$ slope) and a lack of bright AGN
with flux higher than 2$\times$10$^{-11}$ erg s$^{-1}$ cm$^{-2}$.

\end{document}